\documentclass[conference]{IEEEtran}

\usepackage[T1]{fontenc}
\usepackage[bookmarks=false]{hyperref}

\usepackage{amsmath,amssymb,amsfonts}
\usepackage{algorithmic}
\usepackage{graphicx}
\usepackage{tabularx}
\usepackage{textcomp}
\usepackage[dvipsnames,table,xcdraw]{xcolor} 
\usepackage{comment}

\usepackage[binary-units=true]{siunitx}
\usepackage[acronym]{glossaries}
\usepackage{caption}
\usepackage{subcaption}
\usepackage{xspace}
\usepackage{comment}
\usepackage{svg}

\usepackage{multirow}
\usepackage{colortbl}  
\usepackage{enumitem} 
\usepackage{array}
\newcolumntype{L}[1]{>{\raggedright\let\newline\\\arraybackslash\hspace{0pt}}m{#1}}
\newcolumntype{C}[1]{>{\centering\let\newline\\\arraybackslash\hspace{0pt}}m{#1}}
\newcolumntype{R}[1]{>{\raggedleft\let\newline\\\arraybackslash\hspace{0pt}}m{#1}}

\usepackage{rotating}  
\usepackage{wrapfig}
\usepackage{cancel}
\usepackage{amssymb}

\usepackage{pifont}
\definecolor{LightGray}{gray}{0.9}

\newcommand{\cmark}{\textcolor{ForestGreen}{\text{\ding{51}}}}
\newcommand{\xmark}{\textcolor{BrickRed}{\text{\ding{55}}}}
\newcommand{\halfcmark}{\textcolor{Dandelion}{\text{\bcancel{\ding{51}}}}}

\newcommand{\dirquote}[1]{``#1''}

\newcommand{\mytodo}[1]{\textcolor{red}{TODO:~#1}}

\newcommand{\story}[1]{\textcolor{BlueGreen}{$\hookrightarrow$~#1}\\}

\newcommand{\orchName}{Rothniam\xspace}
\newcommand{\miAlgoName}{\textcolor{red}{REMOVE, LEGACY TEXT}\xspace}

\newcommand*\circled[1]{\tikz[baseline=(char.base)]{
		\node[shape=circle,draw,inner sep=0.6pt,font=\sffamily\small] (char) {#1};}}

\newcommand{\singleinfShort}{SI\xspace}
\newcommand{\multiinfShort}{MI\xspace}
\newcommand{\finetuneShort}{FT\xspace}
\newcommand{\shortTPCDS}{TP\xspace}

\makeatletter
\newcommand\notsotiny{\@setfontsize\notsotiny\@vipt\@viipt}
\makeatother

\usepackage[
style=ieee,
backend=biber,
maxcitenames=3,
mincitenames=1,
uniquelist=false,
maxbibnames=999,
maxnames=3, 
minnames=1,
doi=true,
url=false,
isbn=false,
]{biblatex}

\AtBeginBibliography{\small}

\DeclareSourcemap{
	\maps[datatype=bibtex]{
		\map[overwrite=true]{
			\step[fieldset=abstract, null]
		}
	}
}
\DeclareSourcemap{
	\maps[datatype=bibtex]{
		\map[overwrite=true]{
			\step[fieldset=comment, null]
		}
	}
}

\usepackage{enotez}
\usepackage{setspace}
\setenotez{counter-format=symbols}
\setenotez{list-name=Notices}
\DeclareInstance{enotez-list}{custom}{paragraph}
{
	heading = \section*{#1} ,
	notes-sep = 0.2\baselineskip ,
	format = \footnotesize ,
	number = \textsuperscript{#1}
}

\usepackage[%
textwidth=10mm,
textsize=footnotesize
]{todonotes}

\newacronym{vm}{VM}{virtual machine}
\newacronym{ml}{ML}{machine learning}
\newacronym{nn}{NN}{neural network}
\newacronym{llm}{LLM}{Large Language Models}
\newacronym{sla}{SLA}{service level agreement}
\newacronym{lhs}{LHS}{Latin Hypercube Sampling}
\newacronym{ig}{IG}{information gain}
\newacronym{bo}{BO}{Bayesian Optimization}
\newacronym{bbo}{BBO}{black box optimization}
\newacronym{RSSC}{RSSC}{representative sub-space comparison}

\usepackage{scalerel}
\usepackage{tikz}
\usetikzlibrary{svg.path}

\definecolor{orcidlogocol}{HTML}{A6CE39}
\tikzset{
	orcidlogo/.pic={
		\fill[orcidlogocol] svg{M256,128c0,70.7-57.3,128-128,128C57.3,256,0,198.7,0,128C0,57.3,57.3,0,128,0C198.7,0,256,57.3,256,128z};
		\fill[white] svg{M86.3,186.2H70.9V79.1h15.4v48.4V186.2z}
		svg{M108.9,79.1h41.6c39.6,0,57,28.3,57,53.6c0,27.5-21.5,53.6-56.8,53.6h-41.8V79.1z M124.3,172.4h24.5c34.9,0,42.9-26.5,42.9-39.7c0-21.5-13.7-39.7-43.7-39.7h-23.7V172.4z}
		svg{M88.7,56.8c0,5.5-4.5,10.1-10.1,10.1c-5.6,0-10.1-4.6-10.1-10.1c0-5.6,4.5-10.1,10.1-10.1C84.2,46.7,88.7,51.3,88.7,56.8z};
	}
}

\newcommand\orcidicon[1]{\hspace*{1px}\textsuperscript{\footnotesize \href{https://orcid.org/#1}{\mbox{\scalerel*{
					\begin{tikzpicture}[yscale=-1,transform shape]
						\pic{orcidlogo};
				\end{tikzpicture}}{d}}}}}
\title{Efficient and Reuseable Cloud Configuration Search Using Discovery Spaces}

\author{\vspace*{-1\baselineskip}\\\IEEEauthorblockN{
		Michael Johnston\IEEEauthorrefmark{2},
            Burkhard Ringlein\IEEEauthorrefmark{1},
		Christoph Hagleitner\IEEEauthorrefmark{1},
        Alessandro Pomponio\IEEEauthorrefmark{2}, \\
        Vassilis Vassiliadis\IEEEauthorrefmark{2}, 
        Christian Pinto\IEEEauthorrefmark{2},
        and Srikumar Venugopal\IEEEauthorrefmark{2},
	}
		\IEEEauthorblockA{\IEEEauthorrefmark{2}IBM Research Europe -- Dublin, \IEEEauthorrefmark{1}IBM Research Europe -- Zurich
   }\\
}
 
\def\BibTeX{{\rm B\kern-.05em{\sc i\kern-.025em b}\kern-.08em
    T\kern-.1667em\lower.7ex\hbox{E}\kern-.125emX}}
\begin{document}

\maketitle

\begin{abstract}
Finding the optimal set of cloud resources to deploy a given workload at minimal cost while meeting a defined \acrlong*{sla} is an active area of research. Combining tens of parameters applicable across a large selection of compute, storage, and services offered by cloud providers with similar numbers of application-specific parameters leads to configuration spaces with millions of deployment options. 

In this paper, we propose Discovery Space, an abstraction that formalizes the description of workload configuration problems, and exhibits a set of characteristics required for structured, robust and distributed investigations of large search spaces. We describe a concrete implementation of the Discovery Space abstraction and show that it is generalizable across a diverse set of workloads such as \acrlong*{llm} inference and Big Data Analytics. 

We demonstrate that our approach enables safe, transparent sharing of data between executions of best-of-breed optimizers increasing the efficiency of optimal configuration detection in large search spaces. We also demonstrate how Discovery Spaces enable transfer and reuse of knowledge across similar search spaces, enabling configuration search speed-ups of over 90\%.

\end{abstract}

\begin{IEEEkeywords}
Configuration Search, Data abstraction, Optimization, Linked representations, Configuration Management, Knowledge reuse, Data sharing
\end{IEEEkeywords}

\section{Introduction}
\label{sec:introduction}

Cloud customers can choose from a wide range of virtualized and bare-metal node configurations for their application deployments with additional options for memory, networking, and storage~\cite{MahgoubOPTIMUSCLOUDHeterogeneousConfiguration2020}. The set of configurable parameters for the compute resources and the application can be termed as its \emph{configuration space}. Configuration spaces in practice can feature hundreds to thousands of parameter value combinations with varying consequence on time and cost of deployments.

\begin{figure}[t]
\centering{
\includegraphics[width=0.48\textwidth]{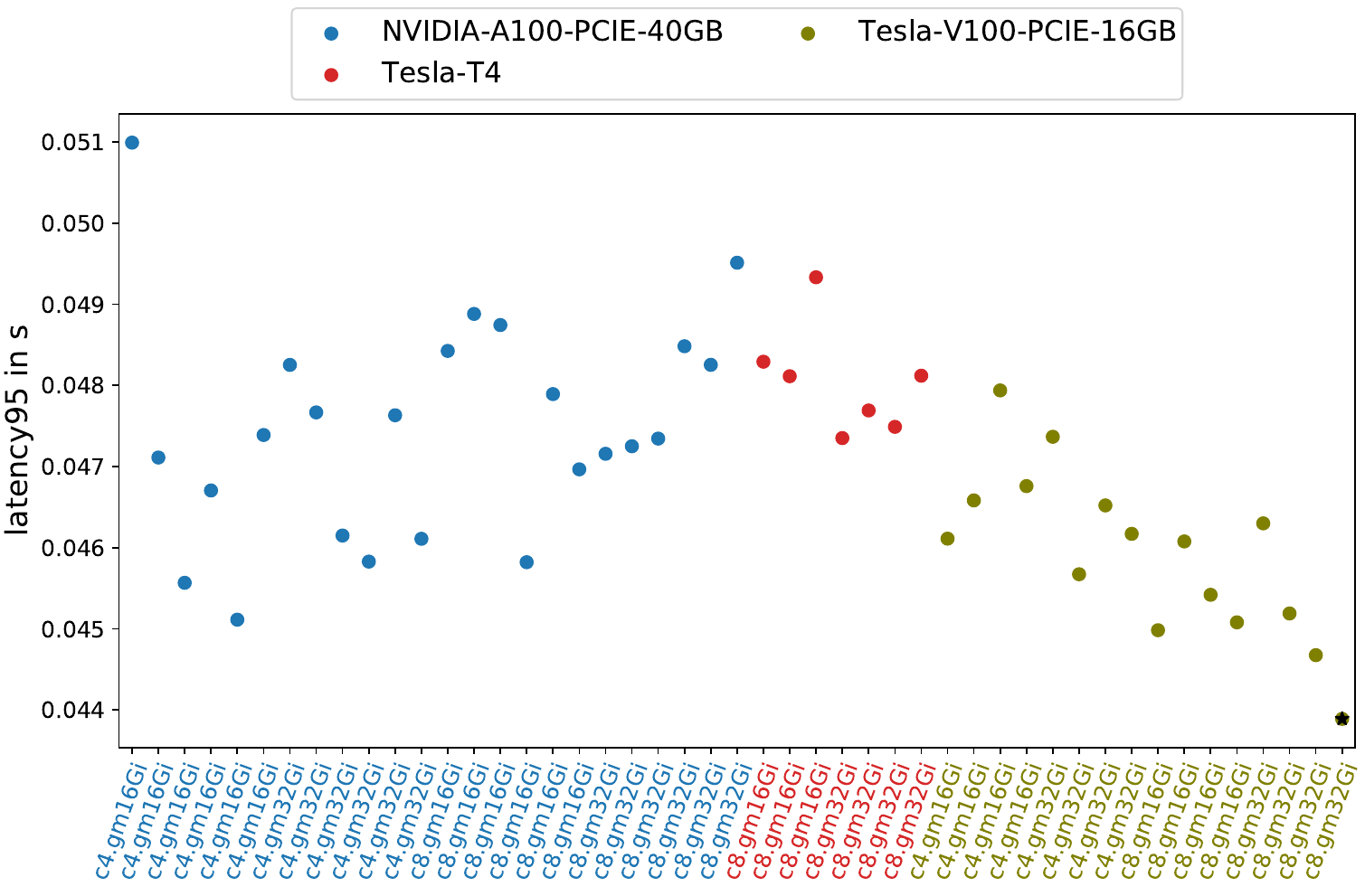}
}
\caption{\label{fig:conf-space-llm}Variation of inference latency ($95^{th}$ percentile) across different configurations for FLAN XL~\cite{wei_finetuned_2022} LLM. The x-axis represents increasing number of CPU cores and GPU memory allocation with 1 GPU of the model represented by the color.}
\end{figure}

Figure~\ref{fig:conf-space-llm} shows an example of such performance variations for an Large Language Model (LLM) inference workload featured in this paper (Section~\ref{sec:evaluation}). For example, 
the configuration with the lowest value for $95^{th}$ percentile latency uses a V100 GPU, which is less capable than A100. However, the highest value is within 10\% of the lowest (best) value. Hence, using the cheapest GPU (Tesla-T4) would provide as much utility as the most expensive (A100). This example illustrates that finding the optimal configuration is non-obvious in face of multiple deployment choices.

Hence, \emph{configuration search} is the problem of finding an optimal resource and application configuration that meets a defined objective for a workload deployment, such as time, cost or energy, among others.  A na\"ive solution to configuration search involves testing of all combinations of parameter values, or the configurations, against the objective. However, this is impossible to accomplish for most practical deployments within a reasonable time or budget. Moreover, some of these parameter combinations may be infeasible due to dependencies which may not be apparent prior to testing~\cite{ZhuBestConfigTappingPerformance2017, MahgoubOPTIMUSCLOUDHeterogeneousConfiguration2020}. 

In practice, configuration search involves sampling the configuration space in conjunction with optimization techniques~\cite{Lazuka2022, ZhuBestConfigTappingPerformance2017, Alipourfard2017, TrotterForecastingGenetic2019}, optionally combined with custom prediction models~\cite{MahgoubOPTIMUSCLOUDHeterogeneousConfiguration2020, Wang2021a, YadwadkarSelectingBestVM2017}. State-of-the-art frameworks such as Vizier~\cite{song_open_2022}, Optuna~\cite{akiba_optuna_2019}, and BOAH~\cite{LindauerBOAHToolSuite2019} provide robust implementations of optimization heuristics along with data structures for managing parameter spaces. Yet, as we will discuss in Sections~\ref{sec:background} and \ref{sec:related-work}, these fall short of meeting challenges such as workload-agnostic search, and being able to reuse results from previous explorations to speed up configuration search. 

\begin{figure*}[t]
	\centering
	\includegraphics[width=0.9\textwidth]{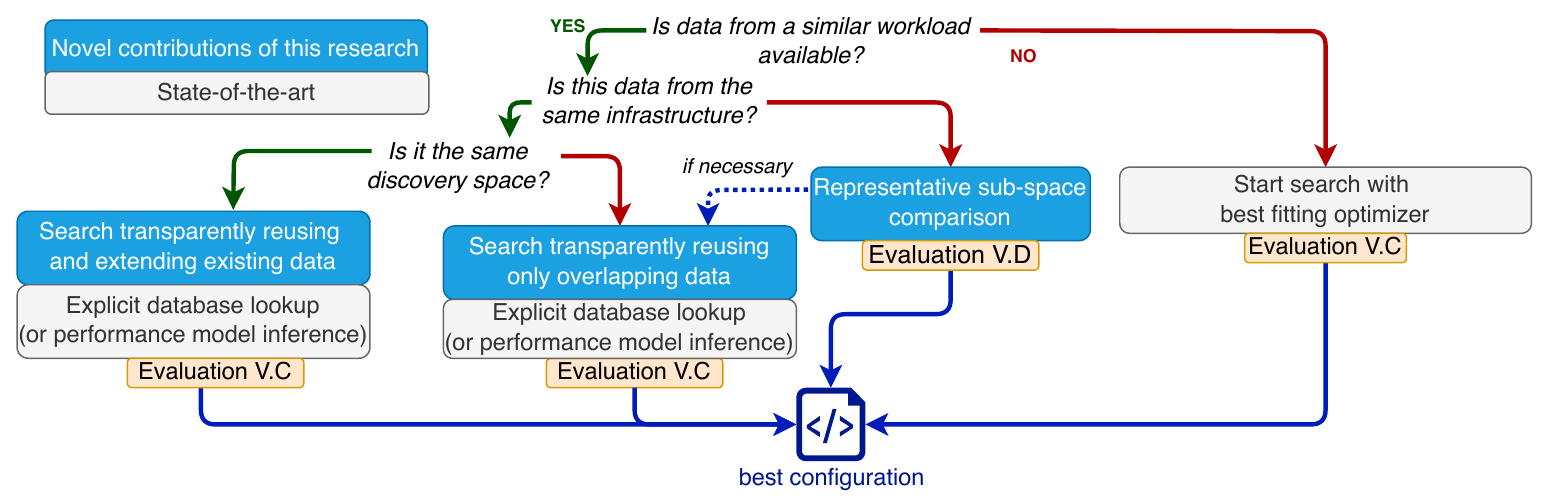}
	\caption{Contributions of this research and the state-of-the-art for a new configuration search, mapped to evaluations.\label{fig:rothniam-help}}
\end{figure*}

The contributions of this research are: 
\begin{enumerate}
	\item A set of required characteristics, called TRACE, for configuration search systems that enables workload-agnostic search and transparent and distributed data-reuse; and a new data-model, targeting configuration search applications, called Discovery Space, that exhibits the TRACE characteristics
 \item An algorithm, called \emph{representative sub-space comparison}, that identifies if data acquired for one configuration space can be transferred to another with similar characteristics, and, if so, performs the transfer. 
	\item A demonstration and evaluation of our Discovery Space implementation on configuration search use-cases in three important domains: Big-Data analytics, \gls*{llm} inference and \gls*{llm} fine-tuning 
\end{enumerate}

The remaining paper is structured as follows: In the next section, we discuss the challenges and set key objectives for ensuring an efficient, reproducible and transferable solution for the configuration search problem. Then, in Section~\ref{sec:methodology}, we derive a set of design principles to meet these objectives, and discuss Discovery Spaces, a data abstraction that is designed around these principles. In Section~\ref{sec:search-impl}, we present our representative subspace comparison method that is based on Discovery Spaces. 
%
Section~\ref{sec:evaluation} presents an in-depth evaluation of our presented methods - Figure \ref{fig:rothniam-help} illustrates how the evaluations map to our objectives.
Finally, we discuss state-of-the-art in detail and present our conclusions.


%

\section{Background and Motivation}
\label{sec:background}

\subsection{Key challenges for configuration search}



\begin{table*}[t]
	\caption{Objectives of this research compared to other workload configuration search and  optimization systems. \xmark, \halfcmark, and \cmark represent none, partially and fully meeting the objective\label{tab:feature-comparison}}

\begin{tabular}{c|ccc|c}
\textbf{Objectives} &
  \begin{tabular}[c]{@{}c@{}}\textbf{Search \& Optimiz.}\\ \cite{Lazuka2022,ZhuBestConfigTappingPerformance2017,Alipourfard2017}\end{tabular} &
  \begin{tabular}[c]{@{}c@{}}\textbf{Prediction Model }\\ \cite{MahgoubOPTIMUSCLOUDHeterogeneousConfiguration2020,Wang2021a,YadwadkarSelectingBestVM2017}\end{tabular} &
  \begin{tabular}[c]{@{}c@{}}\textbf{Black Box Optimizers} \\ \cite{GolovinGoogleVizierService2017,liaw_tune_2018,LindauerBOAHToolSuite2019}\end{tabular} &
  \textbf{\begin{tabular}[c]{@{}c@{}} Discovery Spaces\end{tabular}} \\ \hline
Workload Agnostic             & \xmark     & \xmark     & \cmark & \cmark \\
Multiple Optimization Methods & \halfcmark & \xmark     & \cmark & \cmark \\ \hline
Distributed Sharing           & \xmark     & \halfcmark & \cmark & \cmark \\
Transparent Sharing           & \xmark     & \xmark     & \xmark & \cmark
\end{tabular}

\end{table*}

Table~\ref{tab:feature-comparison} presents a sample of the state-of-the-art for general optimization and configuration search systems juxtaposed with the contributions of this research.


The first challenge is that the featured techniques are customized to specific workloads which limits their applicability. \emph{Workload agnostic optimization} could be implemented through \gls*{bbo} frameworks such as Vizier~\cite{song_open_2022}, Optuna~\cite{akiba_optuna_2019} and BOAH~\cite{LindauerBOAHToolSuite2019} that provide robust and scalable implementations of techniques such as Bayesian optimization. While \gls*{bbo} frameworks have been applied to domains such as hyperparameter tuning of \gls*{llm}, computational chemistry, and finance, these have not found wide application to configuration search. This is in part due to the effort required to map configuration parameters to the formats and techniques used by these tools.

Another challenge is to manage the costs of sampling configuration spaces. One possible solution would be to store and reuse samples gathered from previous explorations of the configuration space. This would not only speed up configuration search through "bootstrapping" but would also amortize the costs of sampling across multiple explorations. In addition, this could accelerate creation of prediction models that require substantial amount of training data which is time-consuming and expensive to collect~\cite{ZhuBestConfigTappingPerformance2017, song_open_2022}. Maintaining provenance of the configuration samples also enables checking if performance models are consistent, thereby enhancing reproducibility~\cite{grayson_benchmark_2024}.

Last but not the least, is the related challenge of managing configuration spaces in dynamic environments. Configuration spaces are impacted by changes in the underlying software and hardware infrastructure, common to cloud environments, which could render existing search solutions obsolete. Repeating the search (regularly) adds time and cost overheads which could be avoided if the validity of the solution can be assessed.

\subsection{Our Objectives}
\label{subsec:motivation}

Our goal was to develop a configuration search framework that would enable efficient and reproducible search of multi-dimensional configuration spaces. Table~\ref{tab:feature-comparison} lists our objectives to achieve this goal. Our \textit{operational objectives }were to support configuration search for any workload with multiple optimization methods using the same framework (\emph{Workload agnostic} and \emph{Multiple Optimization Methods}). 

Our \textit{data-centered objectives} were to identify where existing data is available and transparently use it to save the cost of acquiring it again (\emph{Transparent Sharing}). This enables incremental exploration, where a search can reuse (partial) results from previous searches, as well as aiding reproducibility of the results. Going further than the state-of-the-art, we also aimed at reusing data acquired in one configuration space to inform a search on a different but related configuration space (\emph{Distributed Sharing}).
 
Satisfying both these objectives requires a data model that provides a robust and flexible representation of configuration spaces and their relationship to the data gathered by testing sample configurations. This data model should also abstract the configuration space from the actual optimization techniques used for search.  While recent publications have introduced abstractions for specifying the configuration space~\cite{ZhuBestConfigTappingPerformance2017,Wang2021a}, as yet there is no goal for extending these abstractions to allow reusing and building upon existing experimental data from configuration search studies.

\section{Design and Implementation}
\label{sec:designprinciples}
\label{sec:methodology}

Our aim was to design an abstraction for describing workload configuration spaces, and searches of those spaces, that addresses the operational and data-centered of existing tools (see Table~\ref{tab:feature-comparison}). 
We achieved this by identifying a set of desired characteristics for configuration search systems, called TRACE, and subsequently by deriving a data-model for configuration search, named \textit{ Discovery Space}, which exhibits TRACE characteristics.
The idea is that configuration search applications that adopt the Discovery Space data-model as their core abstraction, will automatically exhibit TRACE characteristics, and hence overcome the identified operational and data-centered limitations.

\subsection{TRACE Characteristics}



We have identified five characteristics of a system we believe are necessary and sufficient for enabling transparent sharing and reuse of data related to search tasks like optimization and exploration. 
These are: Time-Resolved; Reconcilable; Actionable; Common Context; and Encapsulated.
We use the term TRACE to refer to these characteristics, the term \textit{operation} to refer to a task (e.g., an optimization run on a configuration space), and the term \textit{study} to refer to a collection of such operations on the same space. 

\begin{enumerate}[noitemsep, left=0pt]

\item \emph{\textbf{T}ime-Resolved}: The system tracks when and how data is added to a study. If the operations on a study are not time-resolved, it’s not possible to access the time-series of a sampling processes, which are fundamental to uncertainty quantification and understanding the characteristics of the sampling process.
\item \emph{\textbf{R}econcilable}: There is a mechanism that reconciles data in the common context to a study so it appears in a consistent way. If not, operations like optimizations could become inconsistent.
\item \emph{\textbf{A}ctionable}: Access to a study should enable executing measurements which add information to the study. 
\item \emph{\textbf{C}ommon Context}: There is a common storage mechanism and associated schema allowing configurations and associated values that are valid for multiple studies to be shared. Without a common context there is no sharing capability.
\item \emph{\textbf{E}ncapsulated}: A study should encapsulate what configurations and actions are consistent with it, or else it can be contaminated with data unrelated to it. 
\end{enumerate}


The Encapsulated and Reconcilable characteristics motivate a data-model for configuration search that allows operations on the data-model to be stateless. 
This means that users can apply tasks to the data-model at any time and without worrying about side-effects. 
Further, the Actionable characteristic means the actions that change the state of the data-model can be determined from the data-model itself. 


\subsection{Discovery Space Definition}


Our hypothesis is that a data-model built to match a mathematical definition of configuration search will naturally exhibit many of the TRACE characteristics.

\subsubsection{Mathematical Definition}
Configuration search studies have a small number of key properties: a methodology -- the experiments to apply to measure workload performance; a scope -- the resource configuration space under consideration; selection criteria -- which denotes how likely one configuration is compared to another; and the acquired data, which is compatible with the scope and the methodology.

These properties are equivalent to mathematical objects. The scope plus selection criteria form a probability space $(P,\Omega, F)$. 
Here $\Omega$, the sample space, defines the dimensions of the configuration space being explored (e.g. range, value-type) and if they are discrete or continuous. $P$ is a probability measure, often expressed as a probability distribution on each dimension, which governs the selection process. $F$ is the event space, which describes the outcomes of the selection process. In this case, the outcomes are the single configurations, hence $F$ is the elementary event-set and we will omit it for brevity. 

The methodology is equivalent to an Action space, $A$. 
Each element of $A$ is an experiment that can be applied to a configuration to obtain a set of measured property values. 
The Action space defines the measurable properties of a configuration that are of interest e.g. performance properties, and their provenance i.e., the experiment that measured/can measure them. 

The existing data forms a sample set $\{x\}$ and the samples from a particular operation form a time-series, $\{x_{i}, i \in T\}$. Sampling the probability space yields configurations, $\{e\}$. However, we are interested in a configuration, $e$, plus the values given by all the Actions in $A$ applied to it. 
This leads us to the idea of a Discovery Space which we can define as $D = (P,\Omega) \otimes A$ – each element in this space is a configuration, and a set of properties and associated values. Thus, a Discovery Space encapsulates all the data related to configuration search - the space, the samples and the applicable measurement - which aligns to a data-model that enables stateless operations.
It also reinforces that the only way to add data to ${x}$ is via a sampling operation on $D$. 

\subsubsection{Example}
To give a simple example, we can imagine an experiment, \textit{gpu\_flops}, which measures the floating point performance of a given GPU model. This experiment also has a parameter, \textit{batch-size}, controlling the size of the workload tested. We can express an experiment campaign using \textit{gpu-flops} with a Discovery Space, for example: 
\begin{align*}
 D(\Omega, A) = (&'gpu\_model':\{A100,V100\},\\&'batch\_size':\{2,4,8\})  \otimes  \{gpu\_flops\}
\end{align*}
The state of the campaign at any time will be the set $\{x\}$ of points sampled from $D$. 
This might look like: ${(A100, 2 ,0.5 TFLOP), (A100, 4, 0.8TFLOP)}$.
Note, at any point it is known what points have yet to be sampled and how to measure them.

\begin{figure}[t]
\centering{
\includegraphics[width=0.4\textwidth]{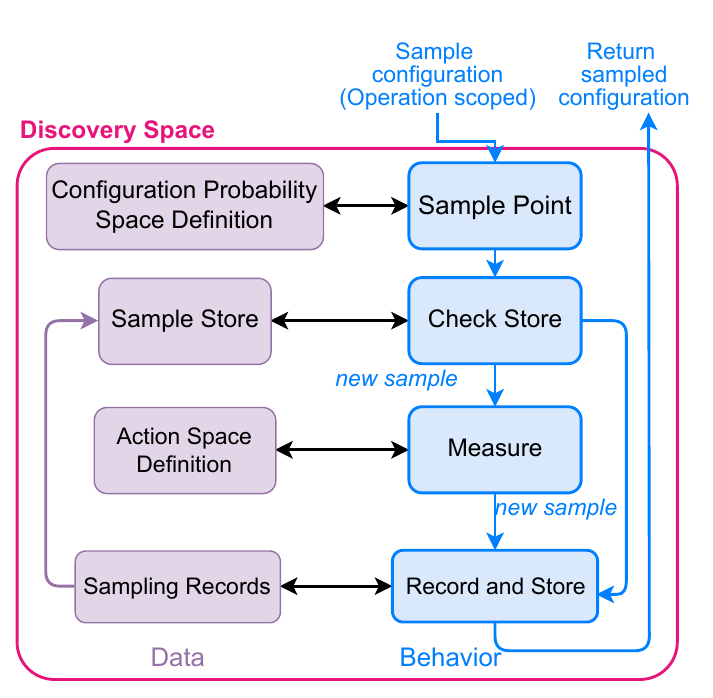}   
}
\caption{A view of a data object representing a Discovery Space. 
\label{fig:discovery-space-stand-alone}}
\end{figure}

\begin{figure}[t]
\centering{
\includegraphics[width=0.4\textwidth]{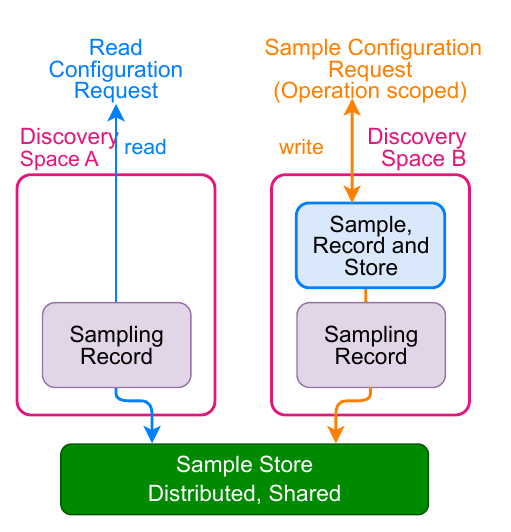}
}
\caption{A view of data sharing between two Discovery Spaces. \label{fig:discovery-space-shared}} 
\end{figure}

\subsection{How a Data Model based on Discovery Space exhibits the TRACE characteristics}
\label{sec:data-model}

This section describes of how the concept of a Discovery Space can be implemented in a configuration search application and how this leads to TRACE characteristics.

\subsubsection{Discovery Space Data Model}

Figure \ref{fig:discovery-space-stand-alone} illustrates the key-features of a Discovery Space data-class implementation. On the left-hand side, it is composed of four data elements which correspond to concepts discussed in the previous section: the configuration probability space, the Action space, the current sample set (sample store), and the recorded operations time-series (sampling records). The sample store and sampling records of all Discovery Space objects are backed by the corresponding storage in the control plane. 

The Discovery Space class provides a method, \textit{sample}, that can write/update samples in the Discovery Space, and a method \textit{read} that can read samples from the Discovery Space. The right-hand side of Figure~\ref{fig:discovery-space-stand-alone} (blue boxes) shows the flow 
when a new configuration is sampled via the \textit{sample} method. In particular, the \textit{Measure} step retrieves information on the associated experiments from the Action Space, executes them, and stores the result in the sampling record and sample store. 

\subsubsection{Enabling the Encapsulated, Actionable and Time-Resolved Characteristics}

The Discovery Space data-model is \textbf{encapsulated} as it defines the allowed measurements and configurations; it is \textbf{actionable} as it contains information on how to obtain measurements; and, by ensuring the sampling sequence of an operation is recorded, it provides \textbf{time-resolution}.  

\begin{figure*}[t]
	\centering
	\includegraphics[width=0.7\textwidth]{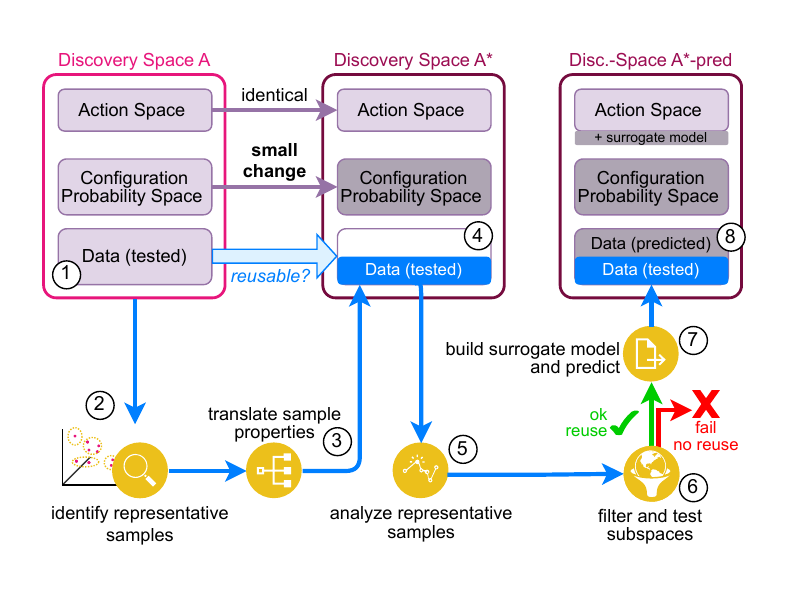}
	\caption{The \textit{representative sub-space comparison} knowledge transfer method. For simplicity this figure shows a change to the configuration space, however it could be to the action space or to both.} 
	\label{fig:rssm-method}
\end{figure*}

\subsubsection{Enabling the Common-Context Characteristic}

Figure \ref{fig:discovery-space-shared} illustrates how two Discovery Spaces objects can share data via a distributed sample store.
The Discovery Space data-model defines a generic schema for specifying configurations, their properties and property values, and the experiments that measured them, that mirrors the mathematical structure of a Discovery Space. This is key for making the Discovery Space data model workload agnostic.  This schema is used to store all sample information in the sample store.  This ensures any data added to the sample store is readable by any operation and hence it enables a \textbf{common context}.  

\subsubsection{Enabling the Reconcilable Characteristic}

The \textbf{reconciliation} mechanism is straightforward - the only way to add data to a Discovery is via its \textit{sample} method. 
As shown in Figure~\ref{fig:discovery-space-shared}, a sampling operation on Discovery Space B triggers a \textit{write/update} to the sampling record and configuration (implies a Measurement was triggered, which is not shown).
In the case that the configuration data written by B is valid for Discovery Space A, a read operation on Discovery Space A will \textbf{not} return the configuration until it is in sampling record of Discovery Space A. That is, even when relevant configuration data is in the common context, it cannot be read via a Discovery Space until a call to that Discovery Spaces \textbf{sample} method generates that configuration. At this point the configurations data will be read from the store rather than being measured again.

\subsubsection{Handling Uncertainty}

In configuration search there may not be single or consistent best configuration and each run of an experiment, even with the same configuration, may give different results.
The Discovery Space model is compatible with this as it allows multiple samples to be associated with each point in the sample space. 
However, a user must decide when running an exploration if they want to only reuse samples associated with a point or reuse and add i.e., also run a new measurement.
For simplicity, here we only consider the scenario where once a sample is added for a point, no further measurements are made.  

\subsection{Implementation}

To evaluate the Discovery Space abstraction we implemented the data-model outlined in Section \ref{sec:data-model}, using a SQL database as a distributed shared sample store (see Figure \ref{fig:discovery-space-shared}). 
We then implemented performance measurement experiments for four different workloads, ensuring they were compatible with this data-model: the TPC-DS benchmark; two LLM inference benchmarks; and a LLM fine-tuning benchmark (see \ref{subsec:environments} and Table \ref{tab:rssm-eval-cfg} for details).
Finally, we wrote a compatibility wrapper enabling Ray Tune, and hence the optimization algorithms it provides, to interact with instances of our Discovery Space data-model. 

\subsection{Summary}

In summary using a Discovery Space data-model for configuration search satisfies the operational and data-centered objectives outlined in Section \ref{subsec:motivation}.
\par{\textit{Workload Agnostic and Multiple Optimization Algorithms}}: In order to be used in an Action Space, all workload specific experiments must be compatible with the schema of the Common Context defined by the Discovery Space data-model. 
Optimization algorithms must adopt the Common Context schema to interact with a Discovery Space, and are decoupled from the workload experiments as they only see the `sample` method of the Discovery Space. 
\par{\textit{Transparent and Distributed Data Sharing}}: The Common Context means all data is stored in the same schema. If the storage is distributed and shared, this data can be accessed by all. Then the  Encapsulated, Reconcilable and Time-Resolved characteristics ensure that multiple operations by multiple users, can understand, use, reuse and add to this data, without adverse affects.  

\section{Accelerating Search via Data Reuse}
\label{sec:search-impl}
\label{subsec:transfer-method}
\label{sec:transfer-method}

A motivation for our distributed-sharing and transparent-sharing objectives is that they can reduce the cost of configuration space explorations via data reuse.
%
TRACEable Discovery Spaces enable by design the reuse of data via \textit{incremental exploration}, where the knowledge of the space is incrementally built up over many individual optimization or exploration runs, and each new run leverages the current knowledge to accelerate itself with respect to prior runs.

However, in workload configuration search it is frequently possible that relevant data from previous searches exists, just for slightly different workloads or infrastructure. 
For example, performance data exists for a GPU-accelerated workload on an existing infrastructure, and administrators want to understand if the performance of this workload will be similar on a new infrastructure with a slightly different GPU model. 
In such cases we have two Discovery Spaces, the \textit{source} and \textit{target}, which are not identical, so samples cannot be directly reused, but that may overlap significantly.
Additionally, dimensions of configuration spaces frequently are categorical, e.g., GPU model. 
The impact of changes to the values of such dimensions can not be easily extrapolated or interpolated from existing data and require sampling the target space.

To address such situations, this work introduces \acrfull*{RSSC}.  
The idea is to identify a few critical points in the source Discovery Space, the \textit{representative sub-space}, and use these to determine if the source knowledge can be-reused. 
In the following sections we describe \acrshort*{RSSC}, including the transfer criteria, and the knowledge transfer methods we have tested.

\subsubsection{Defining Source and Target Discovery Spaces}

A user starts by defining a Discovery Space $A$ that is well understood i.e. has property values for many configurations. See step \circled{1} at the left-hand side of Figure~\ref{fig:rssm-method}.
Then the user defines $A^{*}$ a related Discovery Space with no samples and which is the target of the knowledge transfer.
Note, here we consider the case where $A^{*}$ differs by a small change in the configuration space, however the same process applies if the change is in the Action Space.
The dimensions of the configuration space of $A^{*}$ should be the same as $A$ but the values in each dimension may differ by a mapping, e.g.  for a categorical dimension ``GPU model" mapping the value A100-PCIE in space $A$ to A100-SXM4 in $A^{*}$.
If both configuration spaces are identical then the difference in the two spaces is implicit.



\subsubsection{Representative Sub-Space Identification}

This step involves \textit{identifying representative samples} \circled{2} of Discovery Space $A$.
We do this by clustering the samples in $A$ according to one or more properties of the Action space.
The properties chosen are those the user would like to transfer between the two spaces.
In our current implementation we use silhouette clustering, but other clustering methods are also applicable. 
This ensures the selected points are representative of the global space. 
From each cluster we choose a representative sample, here the centroid of the cluster, and the set of these points is our representative sub-space.

Next, if required, we translate the representative configurations, $\{e\}_{a}$, from  Discovery Space $A$ to their equivalents in Discovery Space $A^{*}$ using the configuration parameter mapping, giving $\{e\}_{a^{*}}$. 
If there is no such mapping then $\{e\}_{a} == \{e\}_{a^{*}}$. 
See step \circled{3} at the bottom of Figure~\ref{fig:rssm-method}.
 
\subsubsection{Transfer Criteria Application}
\label{subsec:transfer_criteria}

In this step, we sample $\{e\}_{a^{*}}$ from  Discovery Space $A^{*}$ \circled{4} and compare the results to $\{e\}_{a}$ using a transfer criteria to understand if the information should be transferred \circled{5}.
In this study the transfer criteria are based statistical properties of a linear regression fit between the absolute values of the metric whose values we would like to transfer from $A$ to $A^{*}$), although other methods could be used. 
Specifically, the transfer criteria are that the linear regression fit has a correlation value ($r$) greater than 0.7, \textbf{and} a p-value below 1\% for the null hypothesis that the slope of the regression is 0.
When both these criteria are met the knowledge is considered transferable. 
If one or neither of these criteria are met the knowledge is not considered transferable.



\subsubsection{Knowledge Transfer}

If the transfer criteria are met, the final step is to transfer the desired property data from Discovery Space $A$.
We do this by creating a surrogate model using the data from the source and target representative sub-spaces (\circled{7}, right-hand side of Figure~\ref{fig:rssm-method}).
The surrogate model used here is the linear relationship identified during the representative sub-space analysis step.

This surrogate model acts as a "predictor" version of the experiments in $A^{*}$ that measure the properties of interest. Since the surrogate model is an additional measurement, adding it to the Action space of $A^{*}$ creates a new Discovery Space, shown in Figure~\ref{fig:rssm-method}), as $A^{*}_{pred}$.
This maintains provenance as it is clear when the source of values is the surrogate model and the action space of $A^{*}$ can still be applied to points to get the real values. 
The final step, shown as \circled{8} in Figure~\ref{fig:rssm-method}), is to use the surrogate model to obtain predictions for the remaining, non-representative sub-space, points in  $A^{*}_{pred}$.

\section{Evaluation}
\label{sec:evaluation}



This section presents an evaluation of our proposed method against the objectives set out in Section~\ref{subsec:motivation} for addressing workload configuration search problems. The structure of this evaluation is also shown in Figure~\ref{fig:rothniam-help}.The aim of the evaluation was to answer the following questions in the context of our test workloads:

\begin{enumerate}[noitemsep, left=0pt]
\item Objective: Multiple optimization algorithms
\begin{enumerate}[noitemsep, left=0pt]
\item Is having access to multiple optimization algorithms beneficial? 

\end{enumerate}

\item Objective: Distributed and Transparent Sharing
\begin{enumerate}[noitemsep, left=0pt]
\item Does the ability to reuse data reduce the cost of finding the optimal configuration? 
\item Can we transfer established knowledge from one search space to another? 
\end{enumerate}
\end{enumerate}

\subsection{Workloads and Environments}
\label{subsec:environments}

\subsubsection{Workloads and Tests}
\label{subsec:sparkworkload}
\label{subsec:llmworkload}


\begin{table*}[t]
\centering
\caption{Evaluation Workloads}\label{tab:workload-specifications}
{	
		\footnotesize
	\begin{tabular}{C{0.1\textwidth}|L{0.35\textwidth}|L{0.35\textwidth}}
		Name & Description & Test and Metric\\ \hline
		TPCDS (\shortTPCDS) & The TPC-DS bench\-mark \cite{AwssamplesEmroneksbenchmark2024} simulates a decision support system 
        running queries over a large amount of data. This workload featured a Apache Spark~\cite{zaharia_spark_2010} implementation of TPC-DS and used a Kubernetes cluster. \emph{Note: only used for optimization tests.} 
		  &  The test consists of 99 queries in total, using a database scale of 100 Gigabytes. The metric is the wall clock execution time. \\ \hline
		\begin{tabular}[c]{@{}c@{}}SINGLE-\\INF (\singleinfShort) \end{tabular} & 

  Tests the latency for inference requests using the Text Generation Inference (TGI)~\cite{tgis:github} and serving the FLAN XL~\cite{wei_finetuned_2022} \gls*{llm} for a single user.
           & 
              $100$ instances of the same inference query are sent sequentially to the \gls*{llm}, mimicking a single user sending requests. The metric is the $95^{th}$ percentile of the request latency's.\\ \hline
        \begin{tabular}[c]{@{}c@{}}MULTI-\\INF (\multiinfShort) \end{tabular}& Tests the latency for inference requests to an LLM served using TGI with a variable numbers of concurrent users
            &  Tests concurrent users (1 to 64) sending requests to the inference server for a total of 10 seconds. The metric is the mean request latency. \\ \hline
        \begin{tabular}[c]{@{}c@{}}FINE-\\TUNE (\finetuneShort) \end{tabular}& Benchmarks the time taken for executing 1 epoch of full fine-tuning using HuggingFace \texttt{accelerate} library with Fully Sharded Data Partitioning (FSDP). \emph{Note: only used for knowledge transfer tests.} 
         &  Tests fine-tuning datasets containing 4096 samples and a set number of tokens per sample. The metric is the total number of input tokens processed per second. 
         \\ \hline
		
	\end{tabular}
 }
\end{table*}

\label{subsec:spaces}

\begin{table*}[tb]

\begin{minipage}{0.48\linewidth}
\centering
\caption{Optimization Tests \label{tab:opt-eval-cfg}}
{
		\footnotesize
\begin{tabular}{l|c|l}
\textbf{Name} &
  \textbf{\begin{tabular}[c]{@{}c@{}}Space\\ Size\end{tabular}} &
  \multicolumn{1}{c}{\textbf{\begin{tabular}[c]{@{}c@{}}Configuration Space \\ (Parameters and Value Sets)\end{tabular}}} \\ \hline
\shortTPCDS-OPT &
  128 &
  \begin{tabular}[c]{@{}l@{}}Executor Instances=12, 14, 16, 18, 20, 22\\ Cores Per Exec.=1, 2, 4, 8\\ Mem.(GB)/Exec.=1, 2, 4, 8, 16\end{tabular} \\ \hline
\singleinfShort-OPT &
  864 &
  \begin{tabular}[c]{@{}l@{}} GPU model=A100-PCIE-40GB, Tesla-T4, \\\#GPU=1,2,4 \\CPU Cores=2,4,8,16\\  V100-PCIE-16GB\\ Container Memory(Gi)=16, 32, 64\\  Max Batch Size=4,24,64,128\\ Max Seq. Length=1024, 2048\end{tabular} \\ \hline
\multiinfShort-OPT &
  2268 &
  \begin{tabular}[c]{@{}l@{}}
  Model Name = Llama2-13B\\Max Batch Size=4,8,16,32,64,128,256\\ Max Batch Weight=19000, 50000, 100000, \\ 1000000, 2000000, 2968750\\ Max Concurrent Requests=64, 128, 320\\ Max New Tokens=512, 1024, 1536\\ Max Sequence Length=1024, 2048, 4096\\ Flash attention=true, false\end{tabular} \\ \hline
\end{tabular}
}
\end{minipage}
\begin{minipage}{0.48\linewidth}
\centering
\caption{Knowledge Transfer Tests \label{tab:rssm-eval-cfg}}
{
	\footnotesize
\begin{tabular}{l|l|l|l}
\textbf{Name} &
   \multicolumn{1}{c|}{\textbf{\begin{tabular}[c]{@{}c@{}}Space\\ Size\end{tabular}}} &
\textbf{Configuration Space} & \textbf{Change} \\ \hline
\multicolumn{1}{l|}{\singleinfShort-TRANS} &
  \multicolumn{1}{l|}{288} &
  \begin{tabular}[c]{@{}l@{}}GPU model=A100-PCIE-40GB\\\#GPU=1,2,4 \\CPU Cores=2,4,8,16\\  Container Memory(Gi)=16, 32, 64\\ Max Batch Size=4,24,64,128\\ Max Seq. Length=1024, 2048\end{tabular} & 
  \begin{tabular}[c]{@{}l@{}} A100-PCIE-40GB \\$\rightarrow$ \\ A100-SXM4-80GB \end{tabular} \\ \hline 
\multicolumn{1}{l|}{\multiinfShort-TRANS} &
  \multicolumn{1}{l|}{48} &
  \begin{tabular}[c]{@{}l@{}}Model Name = Llama2-13B, \\ Granite-13B, Granite-20B\\ Number of CPUs = 1,2,4,8\\ Memory (Gi) = 16, 32, 64, 128\end{tabular} & 
  \begin{tabular}[c]{@{}l@{}} Direct Atch. GPU \\ $\rightarrow$ \\ Network Atch. GPU \end{tabular} \\ \hline
\multicolumn{1}{l|}{\finetuneShort-TRANS} &
  \multicolumn{1}{l|}{56} &
  \begin{tabular}[c]{@{}l@{}}Model Name = llama-7b\\Batch Size = 2,4,8,16, 32,64,128\\ Number of GPUs = 2,4\\ Tokens/Sample = 512, 1024, \\ 2048, 4096\end{tabular} &  
  \begin{tabular}[c]{@{}l@{}} llama-7b \\ $\rightarrow$ \\ mistral-7b  \end{tabular} \\ \hline
\end{tabular}
}
\end{minipage}
\end{table*}

To demonstrate Discovery Space is \emph{workload agnostic}, we performed evaluations using a variety of workloads which are described in Table~\ref{tab:workload-specifications}. The table includes the performance metrics used for evaluations with each workload. For brevity we define short names for the workloads, given in brackets after the full name. 
For all tests we defined workload configuration spaces with  parameters and value ranges considered important from our experience. All these spaces were exhaustively characterized so the best configurations were known, and distributions of the performance metrics were known. Table~\ref{tab:opt-eval-cfg} details the tests used to evaluate optimization. These spaces are large enough so that effective optimization would be preferable over brute-force. Table~\ref{tab:rssm-eval-cfg} details the tests used to evaluate knowledge-transfer via \acrshort*{RSSC}. These cases cover transferring data in cases where resource configuration  (GPU type) and workload configuration (model) change. 
The names of all test-cases incorporate the short-names of the respective workload e.g. SI-OPT, is an optimization test using the SINGLEINF workload.

\subsubsection{Infrastructure}

For test cases based on workloads SINGLEINF, MULTIINF, and FINETUNE, our primary target infrastructure was a bare-metal cluster with 128 core, 512 GB RAM nodes, that provided both direct (PCIE) attached and network attached NVIDIA A100-80GB GPUs. 
We also used a similar cluster that had A100 GPUs directly attached using SXM4 instead of PCIE.  
This was the target for the \multiinfShort-TRANS test. 
For tests based on the TPCDS workload, our target infrastructure was a cloud-hosted Kubernetes cluster with 16 core, 64 GB RAM nodes, provisioned over virtual resources. 

\subsection{Evaluation Methods, Metrics and Baselines}
\label{subsec:evaluationmethods}

\subsubsection{Assessing Optimization Method Performance}
\label{subsec:optimizerperf}

We used three three optimization methods provided via Ray Tune - a classical \gls*{bo} using the \texttt{skopt} library from \texttt{scikit}~\cite{scikit-optimize}; \texttt{Ax}~\cite{ax-platform}, and \texttt{BOHB}~\cite{LindauerBOAHToolSuite2019}- and applied them to the task of finding the best configuration for the optimization tests described in Table \ref{tab:opt-eval-cfg}. The baseline for all tests was the performance of a random walk which is analytically described by the hypergeometric distribution. Each optimizer was used with its off-the-shelf configuration to determine how they behave when used as a black-box.  

We define a "good" configuration as one whose performance is in the $95^{th}$ percentile of the spaces cumulative distribution function (CDF), not including non-deployable points. In the following we denote this metric as \textbf{best\%}. 
To answer question 1a
we performed two evaluations. 
First, for each test-case we executed each optimizer ten times with random starting points, stopping a run if the optimizer did not improve on its current best result in five steps.
The stopping condition is somewhat arbitrary but as it is the same for all optimizers it provides an unbiased point of comparison. 
From this we measured the max and median number of configurations each optimizer samples, and the max and median best\% they achieve. 
Second we evaluated the probability of each optimizer finding a configuration in the $95^{th}$ percentile after sampling N configurations. 
We did this by extending the optimization runs for a given test-case so they all drew the same number of samples, and then calculating the probability of a run having found 1 or more configurations in the target region by that step.
This extension also allows us to examine the behavior of the optimizers after they reached the stopping criteria of the first evaluation. 

To evaluate question 2a for our passive incremental sampling capability, we estimated how much time it would save for the Nth optimization run on a space over the baseline where all measurements in the run are new. To do this we defined a normalized cost for an optimization run as the number of new sample measurements required divided by the total samples (new sample measurements plus samples found in the configuration store). 
This assumes for simplicity that all measurements take equal time. 
For each workload we permuted the order of the optimization runs $100$ times and calculated the average normalized cost for the $i^{th}$ optimization run in the sequence. The runs can be permuted as each is fully independent (following the reconcilable TRACE characteristic). The results for all these evaluations are discussed in section \ref{subsec:ess-eval}.

\subsubsection{Assessing Knowledge Transfer Performance} 
 
To answer questions 2a and 2b for our \acrshort*{RSSC} method, for each test-case in Table~\ref{tab:rssm-eval-cfg} we evaluated for the various scenarios tested: (1) the performance percentile of the best configuration of the predictive model (best\%); (2) the percentage of the actual top 5 best configurations that are in the predictive models top 5 configurations (top5\%); and (3) how well the model can distinguish the ranks of two configurations w.r.t the performance metric (rank resolution). A value of X for the rank-resolution means that the average error of the models predictions is less than the average difference between two configurations separated by X places or greater, and gives a sense of the global fit quality. Note, the knowledge being transferred are the values of the performance metrics of the workloads described in 	Table~\ref{tab:workload-specifications}. To give greater insight into the method we also evaluated transfer of two additional metrics: for \singleinfShort-TRANS the $99^{th}$ percentile of the observed request latency; and for \multiinfShort-TRANS the cumulative latency. 

To evaluate clustering as a method of selecting the representative points for \acrshort*{RSSC} we compared it to two baseline methods: top5 and linspace. For both these methods the points are ranked according to the property to be transferred. Then for top5, the top 5 ranked points are selected, while in linspace, N evenly spaced points over the ranked points are selected. Here N is the number of clusters identified by the clustering method to enable direct comparison. We also evaluated using an optimization run on the target space to identify a representative set of points, bypassing the use of the source space completely. We did this in two ways: (1) by starting the optimizations at the optimal point in the source space (translated as appropriate); (2) by starting at a random point. For a given optimizer run we evaluated the transfer criteria, built linear models, and evaluated the quality metrics. We ran 10 optimizations with each of the three optimizers to ensure we saw average performance. The results of these evaluations are discussed in section \ref{subsec:transfer-eval}.

\subsection{Evaluation of Optimization Methods and Incremental Sampling}
\label{subsec:exhaustive}
\label{subsec:ess-eval}

\subsubsection{Evaluation Aims}
This evaluation has two aims.
First, to evaluate if optimization algorithms perform differently on different workloads. If they do, this is evidence supporting the need for our operational objectives. 
Second, to evaluate if multiple optimization runs from a given method are required to get good results and if sharing data between such these runs would reduce search cost. If they do, this is evidence supporting our data-centered objectives in general and the incremental sampling capability, made possible by the Discovery Space model, specifically.

\subsubsection{TRACE Characteristics Used}
The use of different optimization algorithms is made possible by the Common Context characteristic.
Recording of multiple optimization runs is made possible by the Time-Resolved characteristic.
Distributed incremental sampling leverages the Actionable, Reconcilable, Common Context and Encapsulated characteristics

\subsubsection{Optimizer Performance Evaluation}

\begin{table}[t]
	\centering
	\caption{Summary of the number of trials and performance observed across the 10 individual runs of each optimizer on each workload.
		}
	\label{tab:eval-ess-all}
	{
		\footnotesize
\begin{tabular}{c|crrrr}
\textbf{Test Case} &
  \textbf{\begin{tabular}[c]{@{}c@{}} Optimiz. \\algorithm\end{tabular}} &
  \multicolumn{1}{c}{\textbf{\begin{tabular}[c]{@{}c@{}}max\\ trials\end{tabular}}} &
  \multicolumn{1}{c}{\textbf{\begin{tabular}[c]{@{}c@{}}median\\ trials\end{tabular}}} &
  \multicolumn{1}{c}{\textbf{best \%}} &
  \multicolumn{1}{c}{\textbf{median \%}} \\ \hline
\multirow{3}{*}{\shortTPCDS-OPT}  & Ax          & 13 & 11   & 100  & 94.2 \\
                            & BO  & 14 & 11   & 99.2 & 91.7 \\
                            & BOHB        & 14 & 11.5 & 99.2 & 94.2 \\ \hline

\multirow{3}{*}{\singleinfShort-OPT} & AX & 19 & 15   & 99.7 & 95.2 \\
                           & BO   & 15 & 12   & 99.7 & 97.0 \\
                           & BOHB         & 16 & 14   & 100 & 93.0 \\ \hline

\multirow{3}{*}{\multiinfShort-OPT}   & AX  & 68 & 55.5 & 99.7  & 91.9 \\
                           & BO     & 79 & 39   & 100   & 92.2 \\
                           & BOHB           & 73 & 41   & 99.8  & 95.5 \\ \hline
\end{tabular}
}
\end{table}

\begin{figure}[t]

\centering
\begin{subfigure}{.4\textwidth}

    \includegraphics[width=\linewidth]{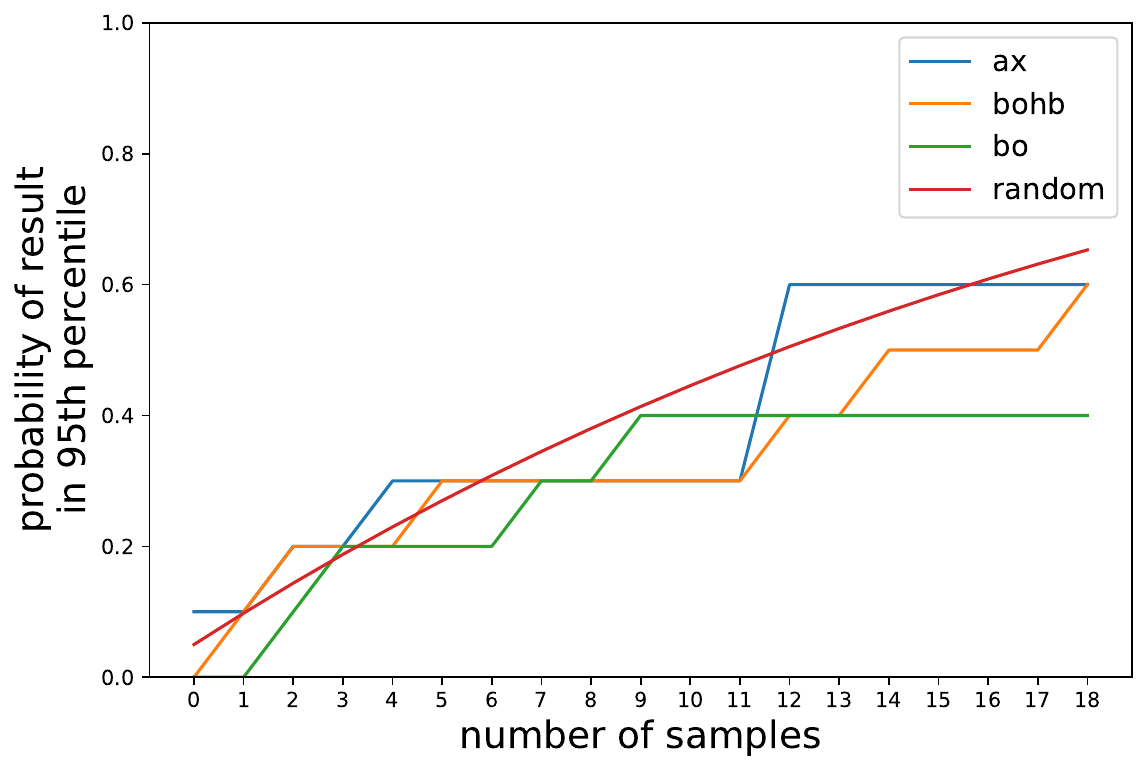}
\end{subfigure}
\begin{subfigure}{.4\textwidth}
    \includegraphics[width=\linewidth]{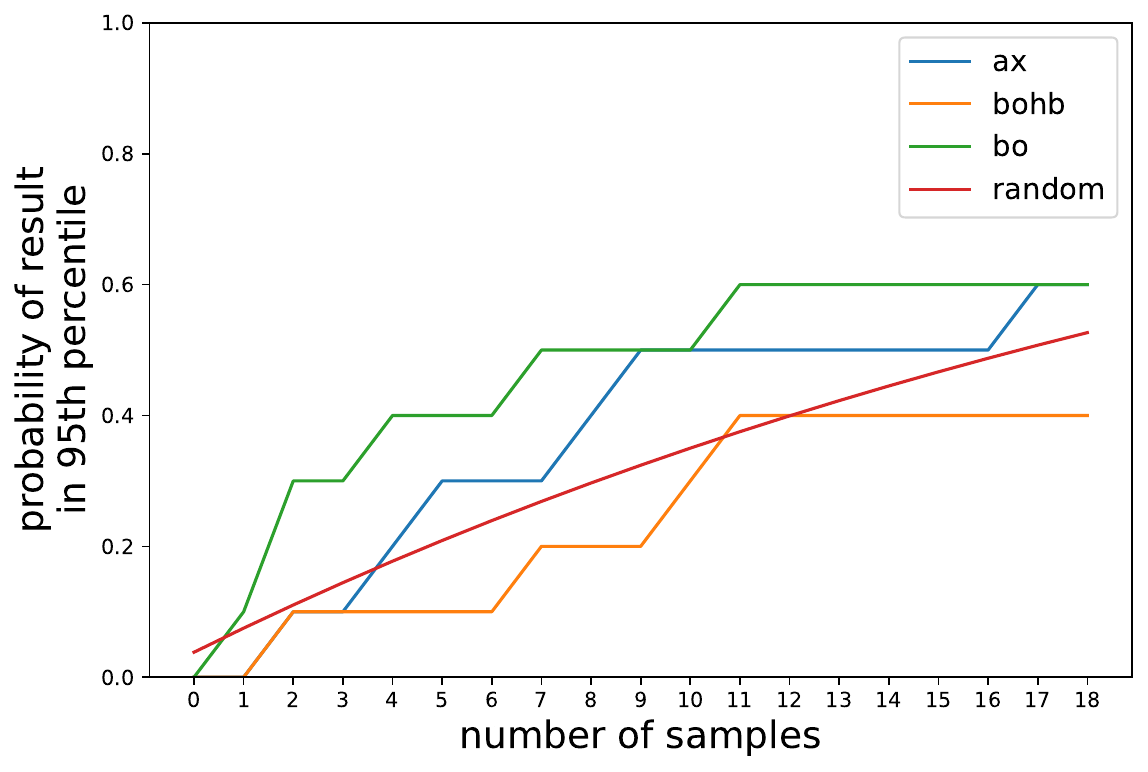}
\end{subfigure}
\begin{subfigure}{.4\textwidth}
    \includegraphics[width=\linewidth]{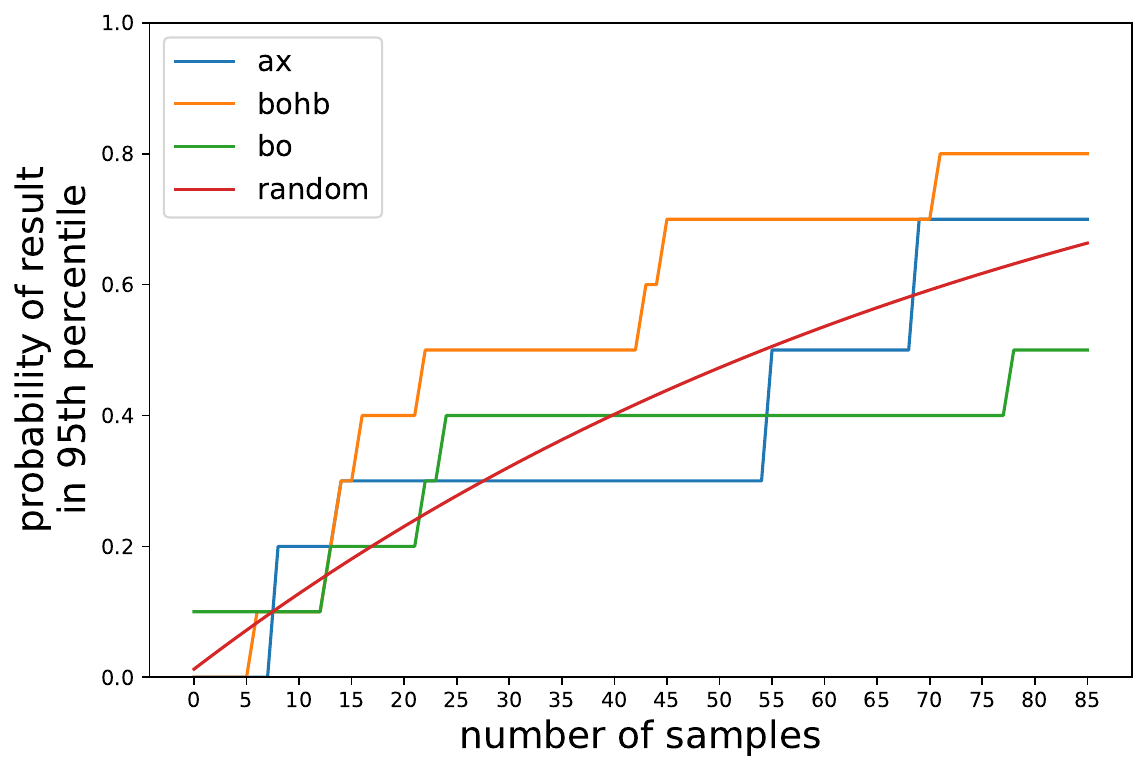}
\end{subfigure}

\caption{Comparison of optimizer performance on the \shortTPCDS-OPT (left), \singleinfShort-OPT (middle) and \multiinfShort-OPT (right) tests for off-the-shelf configurations of optimization algorithms against random sampling (\emph{random}). The plots show the probability of finding one or more configurations in the $95^{th}$ percentile after drawing a given number of samples in a single optimization run.
}
\label{fig:percentbest}
\end{figure}

%
%

Here, we examine the results for the three optimizers on optimization tests described in Table \ref{tab:opt-eval-cfg}. The evaluation methodology and baselines are described in Section \ref{subsec:optimizerperf}. Table~\ref{tab:eval-ess-all} shows if, and how quickly, the optimizers can find configurations in the target-zone for each test-case. From this we can see that for all test-cases all optimizers found in at least one execution the best possible configuration or one very close to it ($99^{th}$ percentile). Further all optimizers halt after sampling less than 2\% of the total search space (median trials divided by space-size, see Table~\ref{tab:opt-eval-cfg}). However, the median optimizer performance varies across test-cases and optimizers. Optimization performs best on \singleinfShort-OPT with the median best\% of BO reaching the $97^{th}$ percentile, while the median performance for \shortTPCDS-OPT and \multiinfShort-OPT are similar, dropping a few percentiles overall. It is noticeable that across the three-tests, three different optimizers produced the best-performance.

Figure \ref{fig:percentbest} sheds more light on the optimizers performance across the test-cases. From this we can see that for \shortTPCDS-OPT the performance of all optimizers is indistinguishable from random-walk. 
However, for \singleinfShort-OPT we see that one optimizer, BO, clearly outperforms random-walk by up to 3x from early in sampling while Ax marginally improves on it. 
For \multiinfShort-OPT, BOHB is clearly better than the others, outperforming random walk by up to 2x, in contrast to Ax and BO whose behaviour is similar to random-walk. These experiments show that state-of-art optimization algorithms can find results close to the absolute minima for our workloads and can significantly out perform random-walk. However, its clear that different algorithms work better in different cases, and that for many cases they performance is equivalent to random-walk. 
This motivates the need for \emph{multiple optimization algorithms}, echoing the results of Bilal, et. al~\cite{bilal_best_2020}.

\begin{figure}[t]
    \includegraphics[width=\linewidth]{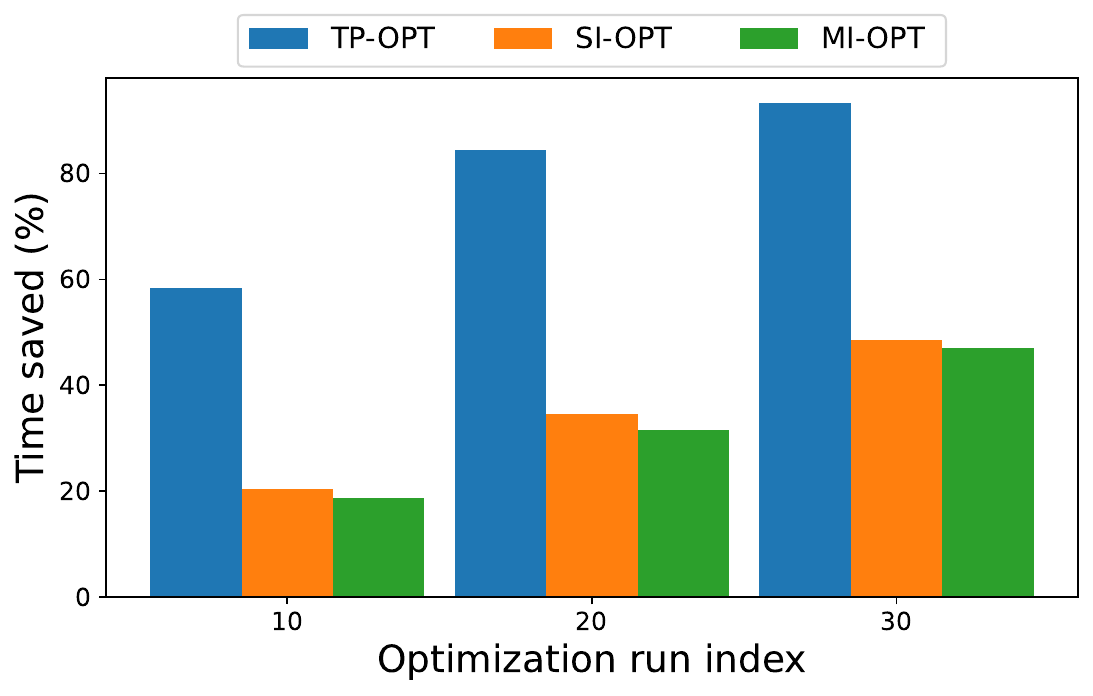}
    \vspace*{-2em}
\caption{Comparison of the impact of re-using existing configuration samples via our shared configuration store for the optimization tests (incremental sampling).
}
\label{fig:incremental}
\vspace*{-1em}
\end{figure}

\subsubsection{Incremental Sampling Evaluation}

These results also demonstrate that without prior knowledge multiple runs will be required to give a significant chance of finding configurations close to the minimum, motivating the need to share and reuse sampling data. 
As mentioned previously (Section \ref{sec:transfer-method}), passive incremental sampling is a method for sharing and reuse that comes "for-free" with Discovery Spaces. 
Figure~\ref{fig:incremental} illustrates the impact of passive incremental sampling feature for the optimization test-cases. The scenario simulated is of multiple researchers sequentially and independently running optimizations with different algorithms on the same Discovery Space. The figure shows the \% time saved due to passive incremental sampling after 10, 20 and 30 runs of average length, w.r.t the baseline of not having this feature. For \shortTPCDS-OPT by the $10^{th}$ run reusing samples saves 60\% of the base optimization runtime. The larger spaces also show considerable savings with 20\% at 10 runs to over 40\% by 30 runs. Interestingly, the progression of savings indicate that the optimizers preferentially sample specific sub-spaces, repeatedly visiting configurations there, while essentially randomly sampling points outside them. For example for \singleinfShort-OPT, the savings are $20\%$ at 10 runs, $34\%$ at 20 ($+14\%$) and $44\%$ at 30 ($+10\%)$. This can be interpreted as the favoured sub-space being sampled quickly giving an immediate reuse boost, with subsequently reuse mostly deriving from random chance and hence the growth in reuse slowing.

\subsection{Evaluation of Representative Subspace Comparison}
\label{subsec:transfer-eval}

\subsubsection{Evaluation Aims}
The aim is to evaluate if \acrshort*{RSSC} can be used to transfer knowledge from one search to another and that this data reuse reduces the cost of finding optimal configurations. 
If it does, this is evidence supporting our data-centered objectives in general and the \acrshort*{RSSC} method in particular. 

\subsubsection{TRACE Characteristics Used}
The  \acrshort*{RSSC} method leverages the Actionable, Reconcilable, Common Context and Encapsulated characteristics

\subsubsection{Evaluation}

We tested \acrshort*{RSSC} using the three transfer tests described in Table \ref{tab:rssm-eval-cfg}. The evaluation methodology and baselines are described in Section \ref{subsec:optimizerperf}. The transfer criteria are detailed in Section \ref{subsec:transfer_criteria}. Table \ref{table:eval-rep-trans-results} shows the transfer and quality metric values for our clustering point-selection and the top5 baseline. We omit the linspace baseline as we found its behavior to be almost identical to clustering. We discuss the advantages of clustering over linspace below. 

\begin{table*}[ht]
	\centering
	\caption{Knowledge Transfer Results - transfer and prediction quality metric values for representative space transfer with the clustering and top5 point selection methods. \cmark indicates that the transfer metric values meet the criteria \xmark indicates the opposite (see Section~\ref{sec:transfer-method}). \halfcmark denotes a situation where one of the two transfer criteria are met. \%savings is time-saved over a brute force evaluation of all configurations. When the predictions should not be used this value is marked as N/A. 
 } 
\label{table:eval-rep-trans-results}
{
\footnotesize

\begin{tabular}{l|l|ll|lll|lll|l}
\textbf{test case} &
  \textbf{metric} &
   \textbf{method} &
 \textbf{\begin{tabular}[c]{@{}c@{}}points\\ selected\end{tabular}} &
  \textbf{r} &
  \textbf{p-value} &
  \textbf{transfer?} &
  \textbf{best\%} &
  \textbf{top5\%} &
 \textbf{\begin{tabular}[c]{@{}c@{}}rank\\ resolution\end{tabular} } &
  \textbf{\%savings}\\ \hline
\multirow{2}{*}{\finetuneShort-TRANS} &
  \multirow{2}{*}{tokens/second} &

  clustering &
  8 &
  0.99 &
  9E-07 &
  \cmark &
  100\% &
  100\% &
  2 &
  86\% \\
 &
   &
  top5 &
  5 &
  0.98 &
  0.003 &
  \cmark &
  100\% &
  100\% &
  2 &
  92\% \\
 \hline
\multirow{4}{*}{\multiinfShort-TRANS} &
  \multirow{2}{*}{cum latency (ms)} &
  clustering &
  4 &
  1.0 &
  0.00014 &
  \cmark &
 80.4\%  &
 60\%  &
  1 &
  92\% \\
 &
   &
  top5 &
  5 &
  -0.025 &
  0.97 &
  \xmark &
  6.5\% &
  0\%  &
  25 &
  N/A \\
 \cline{2-11} 
 &
  \multirow{2}{*}{mean latency (ms)} &

  clustering &
  4 &
  0.98 &
  0.017 &
  \halfcmark &
  95.7\%  &
  60\%  &
  3 &
  92\% \\
 &
   &
  top5 &
  5 &
  0.99 &
  0.0019 &
  \cmark &
  95.7\%  &
  60\%  &
  1 &
  89\% \\
 \hline
\multirow{4}{*}{\singleinfShort-TRANS} &
  \multirow{2}{*}{latency95 (ms)} &
  clustering &
  22 &
  0.29 &
  0.19 &
  \xmark &
  94.8\% &
  0  &
  93 &
  N/A \\
 &
   &
  top5 &
  5 &
  0.72 &
  0.17 &
  \halfcmark &
   94.8\% &
  0\% &
  288 &
  N/A \\
 \cline{2-11} 
 &
  \multirow{2}{*}{latency99 (ms)} &

  clustering &
  33 &
  0.063 &
  0.73 &
  \xmark &
  81\% &
  0\%  &
  92 &
  N/A \\
 &
   &
  top5 &
  5 &
  0.17 &
  0.79 &
  \xmark &
  81\% &
  0\%  &
  237 &
  N/A \\
\hline
\end{tabular}


}
\end{table*}

Our results show that the source and target spaces for \finetuneShort-TRANS and \multiinfShort-TRANS have strong linear relationships, while in the \singleinfShort-TRANS case it is weak. This demonstrates that apparently small changes (PCIE to SXM4), can lead to large perturbations in behavior. Using \acrshort*{RSSC} first can uncover and exploit these strong relationships and quickly flag when they don't exist. 

When \acrshort*{RSSC} uses global point-selection methods (clustering, linspace) the go/no-go transfer criteria map strongly to the resulting quality of linear predictions. This is perhaps not surprising, as when there is a strong linear relationship one expects a method using linear relationships as a criteria and predictor to work well. When the criteria indicate transfer is possible the quality of the predictive model is high e.g. identifying the best or close-to-best configuration, 60\% or more of the top5 and rank resolution of 1 or 2, leading to potentials savings of up to 92\% over the time that would be required for brute-force exploration. In contrast, using a local method like top5 can sometimes give false positives or negatives e.g. cum-latency in \multiinfShort-TRANS and latency95 in \singleinfShort-TRANS. This is because the fit to the behavior around the local minimum may not reflect the global behavior, either showing a linear relationship that is not there globally or vice-versa.

For our optimization baselines we found for \finetuneShort-TRANS and \multiinfShort-TRANS that optimizations, starting at optimal or random points, behave similarly to each other, and \acrshort*{RSSC}, in terms of identifying transfer potential and the resulting quality of the models. 
However, since the optimizers are searching for the minimum they can sample up to 2x more points than \acrshort*{RSSC} for these test-cases e.g. 9 versus 4 for \finetuneShort-TRANS. 
We also found that starting an optimization run from the optimal does not reduce the length of the run, which we attribute to the nature of these gradient-less algorithms. 
Unexpectedly, we found that for \singleinfShort-TRANS there is a strong difference in the behavior of the Ax and BOHB optimizers when starting from the known best point. These runs frequently selected points (25-40\% of the tested runs)  which lead to passing the transfer criteria but that ultimately have the same poor predictive quality for this test-case as the \acrshort*{RSSC} methods shown in Table \ref{table:eval-rep-trans-results}.

In summary \acrshort*{RSSC} provides a way to quickly identify if there is, and is not, strong linear relationships between spaces that can be leveraged to transfer knowledge and save time. Point selection methods that consider the whole space, like clustering and linspace, are preferred. While clustering and linspace perform similarly, clustering has the advantage of not requiring a prior selection of the number of points in the representative set, and also can be used with multiple properties. 
The representative sub-space can also be built via optimization runs on the target space. This approach can be considered if \acrshort*{RSSC} identifies more clusters than a typical optimization run on a space would sample. However, from our results, these runs should not be seeded with information from the source space as this can lead to false positives and negatives.

\section{Related Work}
\label{sec:related-work}

Cloud providers offer a variety of resources with different capabilities at different price points. The complexity of making this choice has spurred interest in the area of configuration search and optimization~\cite{Alipourfard2017, YadwadkarSelectingBestVM2017, Klimovic2018, MahgoubOPTIMUSCLOUDHeterogeneousConfiguration2020}. Configuring data analytics applications and cloud workloads has been one of the focal areas~\cite{Venkataraman2016, Alipourfard2017, hsu_arrow_2018, HuangResourceConfigurationTuning2022, krishna_conex_2022, song_spark-based_2021}. Herodotou, et. al~\cite{herodotou_survey_2021} present a survey of the techniques for these workloads, including Apache Spark jobs~\cite{zaharia_spark_2010}. The configuration of inference services is an area of interest~\cite{zhang_mark_2019} too, due to the configuration spaces featuring diverse application requirements, complex hardware choices, and the presence of model variants~\cite{romero_infaas_2021, Wang2021a, Wu2022}. 

Research in configuration search and optimization has relied on algorithmic techniques such as Bayesian optimization (BO)~\cite{Venkataraman2016, Alipourfard2017, Wang2021a, HuangResourceConfigurationTuning2022, Wu2022, LazukaXCloudServingAutomatedML2023, ZhuBestConfigTappingPerformance2017, bilal_best_2020}, 
genetic~\cite{TrotterForecastingGenetic2019, MahgoubOPTIMUSCLOUDHeterogeneousConfiguration2020}, and multi-arm bandit algorithms~\cite{Lazuka2022}. These have also been used in hyper-parameter optimisation of machine learning models. In particular, both problem areas share many characteristics such as unknown optimization surfaces and large parameter spaces with discrete variables. This motivates the application of BO, and its variants such as Sequential Model-Based Optimization~\cite{bergstra_algorithms_2011}. 

However, even though BBOs such as Vizier~\cite{GolovinGoogleVizierService2017}, BOHB~\cite{falkner_bohb_2018}, Ray Tune~\cite{liaw_tune_2018}, and SageMaker~\cite{liberty_elastic_2020} feature robust implementations of BO, to our knowledge, there have been no application of these frameworks to configuration search. In this paper, our implementation relied on Ray Tune~\cite{liaw_tune_2018} for executing the optimization algorithms using Ray~\cite{MoritzRayDistributedFramework2018} as its distributed computing engine.

Performance modeling has also been used for configuration search, including both black box~\cite{Klimovic2018, YadwadkarSelectingBestVM2017, park_mimir_2023, MahgoubOPTIMUSCLOUDHeterogeneousConfiguration2020} and grey box modeling~\cite{Wu2022, Gao2022} approaches towards gathering data, sometimes in combination with optimization techniques. For example, CherryPick~\cite{Alipourfard2017} and Morphling~\cite{Wang2021a} use performance models to guide the acquisition function of an optimizer while OPTIMUSCLOUD~\cite{MahgoubOPTIMUSCLOUDHeterogeneousConfiguration2020} and PARIS~\cite{YadwadkarSelectingBestVM2017} combine performance models with genetic algorithms and random forests, respectively. While we have not featured these in our evaluation, it is possible to use sampling data of different configurations to build performance models. Therefore, these techniques are complimentary to the aims of our work.

Commonly, these examples from state-of-the-art customize a known algorithm, such as Bayesian Optimization, to meet specific needs of a class of applications. Yet, Bilal, et. al~\cite{bilal_best_2020} show that it is advantageous to have different optimization algorithms for different objectives.  While BestConfig~\cite{ZhuBestConfigTappingPerformance2017} has an extensible architecture that can accommodate different optimization algorithms and different workloads, the result of configuration search in this instance is tied to the specific combination of sampler and optimization algorithm. 

Vizier~\cite{song_open_2022} and Optuna~\cite{akiba_optuna_2019} are example of BBOs with data abstractions (\emph{Studies} and \emph{Trials}) that separate representation of the parameter study space from its operations. These enables features such as multiple optimization algorithms over the same study and transfer of knowledge by using earlier searches as priors. Similarly, our Discovery Space abstraction is agnostic to both \emph{workload} and \emph{optimization} which enabled evaluations as shown in Section~\ref{subsec:ess-eval}, where different optimizers can be applied to the same set of sampling data.

Meeting the data sharing objectives through Discovery Spaces enables the reuse of data as described in Section~\ref{subsec:transfer-method}. Vizier, Optuna and Botorch~\cite{balandat_botorch_2020} are BBOs that support reuse of past studies to seed optimization executions but the selection of sources are the user's responsibility. Discovery Spaces on the other hand, guarantee that only relevant data is used, and this can be done transparently to the user (demonstrated in Section~\ref{sec:transfer-method}). Furthermore, our RSSC technique, is able to automatically predict values for the target space based on identifying a few representative samples. To our knowledge, this is the first instance of the joint application of automatic selection and prediction applied to configuration search and optimization.

\section{Conclusion}


%


This paper began by setting objectives for efficient and workload-agnostic search for optimal configurations in high-dimensional spaces. We presented the TRACE principles and Discovery Space data model to achieve these objectives.
We showed how these principles make configuration search workload-agnostic, by determining the best optimization algorithm for searching the  configuration spaces of three different application workloads: For the MULTIINF workload, BOHB has higher than $80\%$ chance of finding a configuration in the top $95^{th}$ percentile of performance in a single run sampling 3.2\% of the space, while for SINGLEINF and TPCDS the best optimizer is respectively Scikit's \acrlong*{bo} implementation and AX.

We also demonstrated how the TRACE principles enable transparent re-use of results across optimization runs, leading to incremental exploration of configuration spaces and reduction of the cost of exploration by up to $46\%$ for SINGLEINF and MULTIINF workloads and $94\%$ for TPCDS. They also enable distributed knowledge sharing and we presented a method for determining if previous knowledge can be reused for a new application or infrastructure configuration and transferring it if so. Reusing the knowledge of a configuration search for new infrastructure saved $92\%$ or $7$ hours of sampling to determine a good initial configuration. 


In conclusion, the TRACE principles and Discovery Space data-model enable workload agnostic configuration search via multiple search methods along with distributed data-sharing and transparent data-reuse, features that can significantly accelerate configuration search. 
They also enable a flexible modular architecture where it is easy to combine multiple methods e.g., sampling with performance models or using iterative refinement, in novel ways. 
Future work can leverage this capability to realize new methods for identifying relationships between sub-spaces of a source and target space or to identify, and leverage, strong relationships across a set of source spaces that  map to target space in the fewest samples possible.

\printbibliography


\end{document}